\documentclass[%
 reprint,
superscriptaddress,
 amsmath,amssymb,
 aps,
]{revtex4-2}

\usepackage{graphicx}
\usepackage{dcolumn}
\usepackage{bm}
\usepackage{color}
\usepackage{multirow}
\usepackage{booktabs}

\usepackage{color}
\usepackage[colorlinks,bookmarks=false,citecolor=darkblue,linkcolor=red,urlcolor=blue]{hyperref}
\definecolor{darkblue}{rgb}{0,0.02,0.45}

\newcommand{\cmo}{Co$_{2}$Mo$_3$O$_8$}
\newcommand{\fmo}{Fe$_{2}$Mo$_3$O$_8$}

\newcommand{\mmo}{Mn$_{2}$Mo$_3$O$_8$}
\newcommand{\TN}{\ensuremath{T_{\mathrm{N}}}}

\newcommand{\Epara}{$\boldsymbol{E}^{\omega}\parallel c$}
\newcommand{\Eparaa}{$\boldsymbol{E}^{\omega}\parallel a$}

\newcommand{\Hpara}{$\boldsymbol{H}^{\omega}\parallel c$}
\newcommand{\Hparaa}{$\boldsymbol{H}^{\omega}\parallel a$}

\newcommand{\dd}{directional dichroism}

\begin{document}

\title{Confirming the trilinear form of the optical magnetoelectric effect\\ in the polar honeycomb antiferromagnet \cmo{} }
\date{\today}

\author{S.~Reschke}
\affiliation{Experimental Physics V, Center for Electronic
Correlations and Magnetism, Institute for Physics, Augsburg
University, D-86135 Augsburg, Germany}
\affiliation{TOPTICA Photonics AG, Lochhamer Schlag 19, 82166 Gr\"afelfing, Germany}
\author{D. G.~Farkas}
\affiliation{Department of Physics, Budapest University of Technology and Economics, 1111 Budapest, Hungary}
\affiliation{Condensed Matter Research Group of the Hungarian Academy of Sciences, 1111 Budapest, Hungary}
\author{A.~Strini\'{c}}
\author{S.~Ghara}
\affiliation{Experimental Physics V, Center for Electronic
Correlations and Magnetism, Institute for Physics, Augsburg
University, D-86135 Augsburg, Germany}
\author{K.~Guratinder}
\author{O.~Zaharko}
\affiliation{Laboratory for Neutron Scattering and Imaging, Paul Scherrer Institut, CH-5232 Villigen PSI, Switzerland}
\author{L.~Prodan}
\author{V.~Tsurkan}
\affiliation{Experimental Physics V, Center for Electronic
Correlations and Magnetism, Institute for Physics, Augsburg
University, D-86135 Augsburg, Germany} \affiliation{Institute of
Applied Physics, MD-2028~Chi\c{s}in\u{a}u, Republic of Moldova}
\author{D.~Szaller}
\affiliation{Institute of Solid State Physics, TU Wien, 1040 Vienna, Austria}
\author{S.~Bord\'{a}cs}
\affiliation{Department of Physics, Budapest University of Technology and Economics, 1111 Budapest, Hungary}
\author{J.~Deisenhofer}
\author{I.~K\'{e}zsm\'{a}rki}
\affiliation{Experimental Physics V, Center for Electronic
Correlations and Magnetism, Institute for Physics, Augsburg
University, D-86135 Augsburg, Germany}

\date{\today}

\begin{abstract}
Magnetoelectric phenomena are intimately linked to relativistic effects and also require the material to break spatial inversion symmetry and time reversal invariance. Magnetoelectric coupling can substantially affect light-matter interaction and lead to non-reciprocal light propagation. Here, we confirm on a fully experimental basis, without invoking either symmetry-based or material-specific assumptions, that  the optical magnetoelectric effect in materials with non-parallel magnetization ($\boldsymbol{M}$) and electric polarization ($\boldsymbol{P}$) generates a trilinear term in the refractive index, $\delta n\propto\boldsymbol{k}\cdot(\boldsymbol{P}\times\boldsymbol{M})$, where $\boldsymbol{k}$ is the propagation vector of light. Its sharp magnetoelectric resonances, that are simultaneously electric and magnetic dipole active excitations, make \cmo{} an ideal compound to demonstrate this fundamental relation via independent variation of $\boldsymbol{M}$, $\boldsymbol{P}$ and $\boldsymbol{k}$. Remarkably, the material shows almost perfect one-way transparency in moderate magnetic fields at some of the magnetoelectric resonances.

\end{abstract}

\maketitle

\section*{Introduction}

The intense research on magnetoelectric and multiferroic compounds in recent years~\cite{Tokura:2017,Tokura:2018,Kocsis:2019,Weymann:2020,Cano:2021} has revealed a plethora of novel optical phenomena specific to these materials~\cite{Rikken:2002,Jung:2004,Kubota:2004,Pimenov:2006,Saito:2008,Saito:2008a,Kezsmarki:2011,Bordacs:2012,Takahashi:2012,Takahashi:2013,Szaller:2013,Kezsmarki:2014,Szaller:2014,Kuzmenko:2014,Kuzmenko:2015,Kezsmarki:2015,Bordacs:2015,Toyoda:2015,Kurumaji:2017a,Kurumaji:2017b,Iguchi:2017,Tokura:2018,Kocsis:2018,Kuzmenko:2018prl,Yu:2018,Szaller:2019,Kuzmenko:2019,Viirok:2019,Kimura:2020,Yokosuk:2020,Vit:2021,Weymann:2021,Toyoda:2021}. Most of these optical effects can be rooted back to the simultaneous breaking of the time-reversal and the spatial inversion invariance in these compounds  due to their coexisting magnetic and electric orderings~\cite{Jung:2004,Kubota:2004,Pimenov:2006,Saito:2008,Saito:2008a,Kezsmarki:2011,Bordacs:2012,Takahashi:2012,Takahashi:2013,Szaller:2013,Kezsmarki:2014,Szaller:2014,Kuzmenko:2014,Kuzmenko:2015,Kezsmarki:2015,Bordacs:2015,Toyoda:2015,Kurumaji:2017b,Iguchi:2017,Tokura:2018,Cheong:2018,Szaller:2019,Kimura:2020,Yokosuk:2020,Toyoda:2021}.
Perhaps the most exotic optical effect recognized so far in magnetoelectric media is the non-reciprocal directional dichroism \cite{Rikken:2002,Jung:2004,Kubota:2004,Saito:2008,Saito:2008a,Kezsmarki:2011,Bordacs:2012,Takahashi:2013,Szaller:2013,Kezsmarki:2014,Szaller:2014,Kuzmenko:2015,Kezsmarki:2015,Bordacs:2015,Toyoda:2015,Kurumaji:2017b,Iguchi:2017,Kocsis:2018,Yu:2018,Szaller:2019,Viirok:2019,Kimura:2020,Yokosuk:2020,Vit:2021}, when counter-propagating beams with the same initial polarization are transmitted differently through the medium. In terms of light intensity, it is manifested in the so-called directional dichroism~\cite{Brown:1963,Arima:2008}, when the magnitude of light absorption is different for the two beams with opposite ($\pm\mathbf{k}$) propagation vectors. This effect is of relativistic origin and usually considered to be weak~\cite{Krichevtsov:1996,Rikken:2002,Jung:2004,Kubota:2004,Shimada:2006}. However, in some compounds the so-called one-way transparency, i.e.~the maximal directional anisotropy, has been achieved in resonance with magnetoelectric excitations~\cite{Kezsmarki:2014,Kuzmenko:2015,Bordacs:2015,Toyoda:2015,Yu:2018,Viirok:2019}.

The symmetry-breaking via applied electric and magnetic fields is well-known to have profound effects on the light propagation in solids. In fact, apart from the first experimental demonstration of directional dichroism in the exciton resonances of a polar crystal in magnetic field~\cite{Hopfield:1960}, early realizations of the effect were achieved on paraelectric and paramagnetic systems subject to external electric and magnetic fields~\cite{Rikken:2002,Shimada:2006}.
As pointed out by Rikken and coworkers, in materials exposed to perpendicular static electric $\boldsymbol{E}$ and magnetic $\boldsymbol{B}$ fields, the refractive index has a term proportional to $\boldsymbol{k}\cdot(\boldsymbol{E}\times\boldsymbol{B})$, hence, its magnitude is different for light propagation along $\pm\boldsymbol{k}$ \cite{Rikken:2002}. This polarization-independent effect was argued to be of relativistic origin, inherent to every material, since it can be traced back to the usual magnetic linear birefringence/dichroism, also called Cotton-Mouton or Voigt effect, via a Lorentz boost. The magnetic linear birefringence/dichroism, that is quadratic in $B$, describes the difference in the real/imaginary part of the refractive index for light polarizations along and perpendicular to an externally applied magnetic field. In a frame moving with a velocity of $\boldsymbol{v}=c\cdot \boldsymbol{E}\times\boldsymbol{B}/B^2$ with respect to the original frame, there is a static electric field emerging and the optical anisotropy originally $\propto B^2$ is transformed to the directional optical anisotropy $\propto\boldsymbol{k}\cdot(\boldsymbol{E}\times\boldsymbol{B})$.

Even without considering the microscopic origin, one can generally argue on symmetry basis that such a triple-product term can be present in the refractive index. All fundamental interactions, except for the weak interaction, obey separately the time reversal, the spatial inversion and the charge conjugation symmetries. Therefore, if any of these three operations are simultaneously applied to the light field and the material, the measured refractive index should stay invariant. In fact, this triple-product form fulfills this condition, since either of these three operations reverse two vectors of the triple product: Time reversal switches the sign of $\boldsymbol{k}$ and $\boldsymbol{M}$, inversion reverses $\boldsymbol{k}$ and $\boldsymbol{P}$, and charge conjugation reverses $\boldsymbol{P}$ and $\boldsymbol{M}$. Concerning spatial symmetries, the triple product is an invariant scalar not only with respect to inversion but to all symmetry operations in $O(3)$, as expected for light-matter interaction.

In these first proof-of-concept experimental studies, carried out on paraelectric and or paramagnetic materials, the magnitude of the directional dichroism was found to be of the order of a few percents at most \cite{Rikken:2002,Shimada:2006}. Since externally induced and spontaneous built-in fields are equivalent in terms of symmetry, directional dichroism of the form $\delta n \propto \boldsymbol{k}\cdot (\boldsymbol{P}\times\boldsymbol{M})$ was expected to emerge in multiferroic compounds with finite crossed polarization ($P$) and magnetization ($M$). In addition to materials with finite $P$ and $M$, directional optical anisotropy can produce contrast between antiferromagnetic domains~\cite{Brown:1963}, as demonstrated in non-centrosymmetric (antipolar) antiferromagnetic crystals, where the reversal of $\boldsymbol{k}$ is equivalent to the reversal of the antiferromagnetic N\'eel vector~\cite{Kocsis:2018} or to the inversion of the quadrupole moment of the domain~\cite{Kimura:2020}.

While the trilinear form of the directional anisotropy has not been experimentally demonstrated in full extent in multiferroics, the effect was found to be highly amplified by the built-in symmetry breaking fields, in some cases leading to one-way transparency~\cite{Saito:2008a,Kuzmenko:2015,Bordacs:2015,Toyoda:2015,Yu:2018}. Parts of the trilinear expression were verified recently in various multiferroic or magnetoelectric crystals, by showing that $\delta n$ changes sign upon the flipping of $\boldsymbol{M}$~\cite{Rikken:2002,Jung:2004,Kubota:2004,Shimada:2006,Saito:2008a,Kezsmarki:2015,Bordacs:2015,Yu:2018}, $\boldsymbol{k}$~\cite{Toyoda:2015}, $\boldsymbol{M}$  and/or $\boldsymbol{P}$~\cite{Kezsmarki:2011,Takahashi:2012,Takahashi:2013}, $\boldsymbol{M}$  and/or $\boldsymbol{k}$~\cite{Kuzmenko:2015}. The only study investigating the effect of the one-by-one reversal of all the three vectors was carried out on Ba$_2$Mg$_2$Fe$_{12}$O$_{22}$, where the triple-product form of the optical magnetoelectric effect in the microwave region was rigorously demonstrated on experimental basis~\cite{Iguchi:2017}.

\begin{figure}[t!]
\includegraphics[width = \columnwidth]{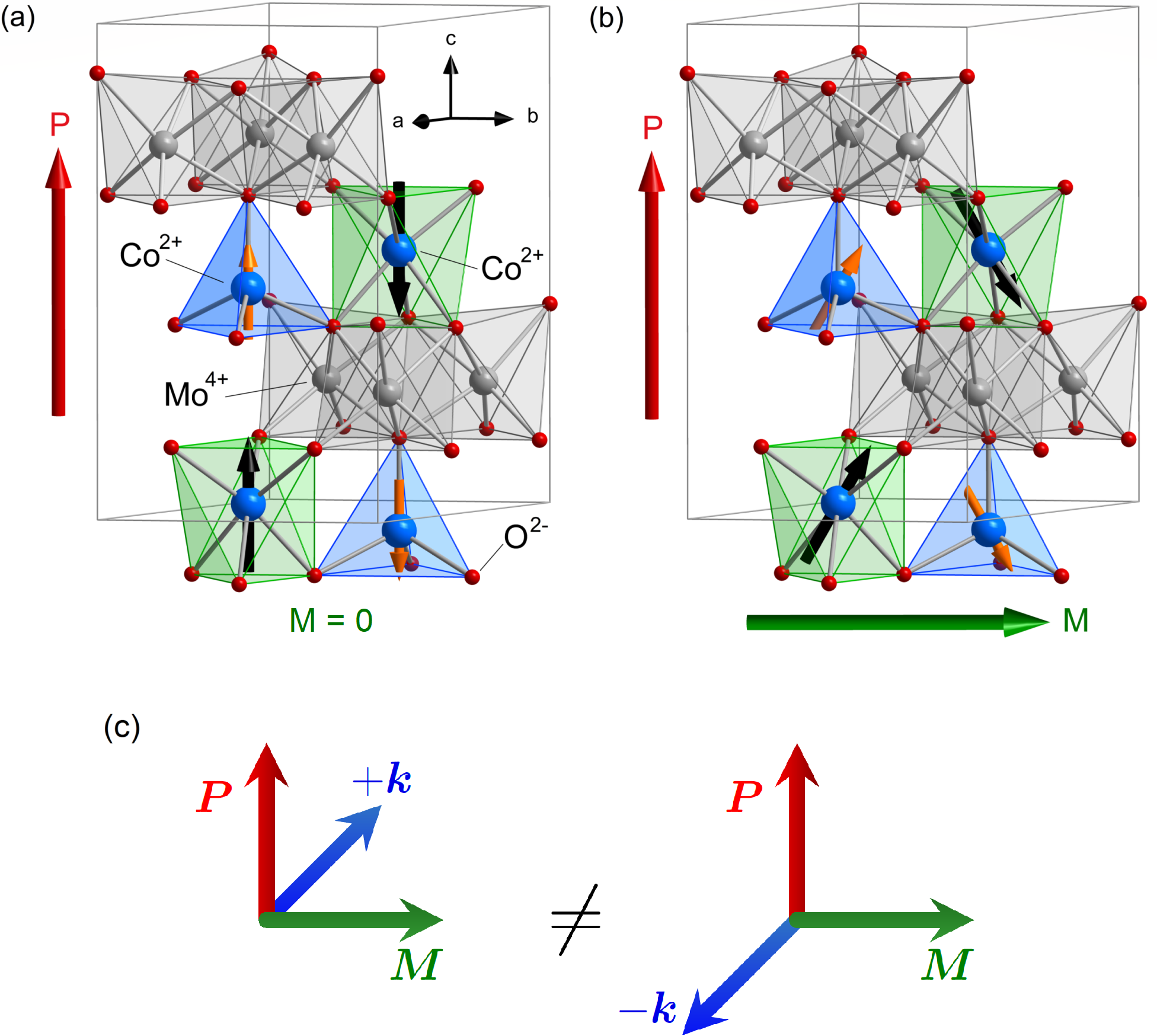}
\caption{\label{fig:Crystal_structure} (a) Crystallographic structure of \cmo{}, with ferroelectric polarization $P$ along the $c$-axis. The magnetic Co$^{2+}$ ions are located in both tetrahedral (blue) and octahedral (green) oxygen coordination. The antiferromagnetic spin arrangement is indicated by the arrows. (b) A magnetization $M$ is induced for magnetic field $\boldsymbol{H}\perp c$  by a canting of the spins. (c) Illustration of the triple product $\boldsymbol{k}\cdot (\boldsymbol{P}\times\boldsymbol{M})$ configurations leading to directional dichroism.}
\end{figure}

Here, we choose the polar easy-axis antiferromagnet \cmo{} \cite{Tang:2019} as a benchmark material to systematically test the triple-product form, $\boldsymbol{k}\cdot (\boldsymbol{P}\times\boldsymbol{M})$, of the directional optical anisotropy. \cmo{} belongs to the family of type-I multiferroic transition-metal molybdenum oxides M$_{2}$Mo$_3$O$_8$  (M = Mn, Fe, Co) with a hexagonal structure in the polar space group $P6_3mc$ \cite{McCarroll:1957,Varret:1972,Bertrand:1975,LePage:1982,Strobel:1982,McAlister:1983,Abe:2010} (see Fig. \ref{fig:Crystal_structure} for the crystal structure). The Co$^{2+}$ ions, located in both octahedrally and tetrahedrally coordinated sites, are responsible for the magnetism, as  the Mo$^{4+}$ ions form non-magnetic trimers \cite{Cotton:1964,Varret:1972,Wang:2015}. Due to its polar crystal structure {\cmo} has a pyroelectric polarization $\boldsymbol{P}$ along the $c$-axis and below $\TN{}=40$\,K it undergoes a transition to a four-sublattice easy-axis collinear antiferromagnetic state with the N\'eel vector $\boldsymbol{L}$ pointing along the $c$-axis [see Fig. \ref{fig:Crystal_structure}(a)] \cite{Bertrand:1975,McAlister:1983,Abe:2010,Tang:2019}. Similarly to other members of the M$_{2}$Mo$_3$O$_8$ family \cite{Kurumaji:2015,Wang:2015,Kurumaji:2017}, the polarization of \cmo{} is also affected by the magnetic ordering, implying a remarkable magnetoelectric coupling \cite{Tang:2019}. When the magnetic field is applied perpendicular to the $c$ axis, a sizable magnetization can be induced via the canting of the spins away from the $c$ axis, as sketched in Fig. \ref{fig:Crystal_structure}(b). This state with perpendicular $\boldsymbol{P}$ and $\boldsymbol{M}$ is expected to exhibit directional anisotropy\cite{Kezsmarki:2011}, with different refractive indices for beams travelling along the $\pm\boldsymbol{k}$ direction, as depicted in \ref{fig:Crystal_structure}(c).

Similarly to the cases of the antiferromagnetic \fmo~\cite{Kurumaji:2017a}, and ferrimagnetic Zn-doped \fmo~\cite{Kurumaji:2017b,Csizi:2020} and \mmo~\cite{Szaller:2020}, in the low-energy range ($<$16\,meV) we observe two strong magnetic excitations in zero field both in the terahertz absorption and the inelastic neutron scattering data. In contrast to the sister compounds, both of these modes in {\cmo} are doubly degenerate and show a V-shape splitting in magnetic fields applied along the $c$ axis, as expected for a four-sublattice easy-axis collinear antiferromagnet. When {\cmo} is magnetized perpendicular to the $c$ axis, i.e. perpendicular to its pyroelectric polarization, these modes exhibit a strong directional dichroism for light beams propagating perpendicular to both $\boldsymbol{P}$ and $\boldsymbol{M}$. We systematically demonstrate via the sequential change of $k$, $P$ and $M$ that the observed directional dichroism corresponds to a trilinear term in the refractive index, $\delta n \propto \boldsymbol{k}\cdot (\boldsymbol{P}\times\boldsymbol{M})$. Besides these dominant features, additional weaker excitations are resolved in the antiferromagnetic state by terahertz (THz) spectroscopy.




\section*{Results}
\noindent
\textbf{Dispersion of the magnetic excitations}\\
According to linear spin wave theory, in a simple collinear antiferromagnet, the number of magnon modes at each wavevector is expected to be equal to the number of magnetic sublattices. Given that {\cmo} realizes a four sublattice collinear antiferromagnetic order below $T_N$, four modes are expected, which form two doublets in zero field.

\begin{figure}[!t]
\includegraphics[width = 1\columnwidth]{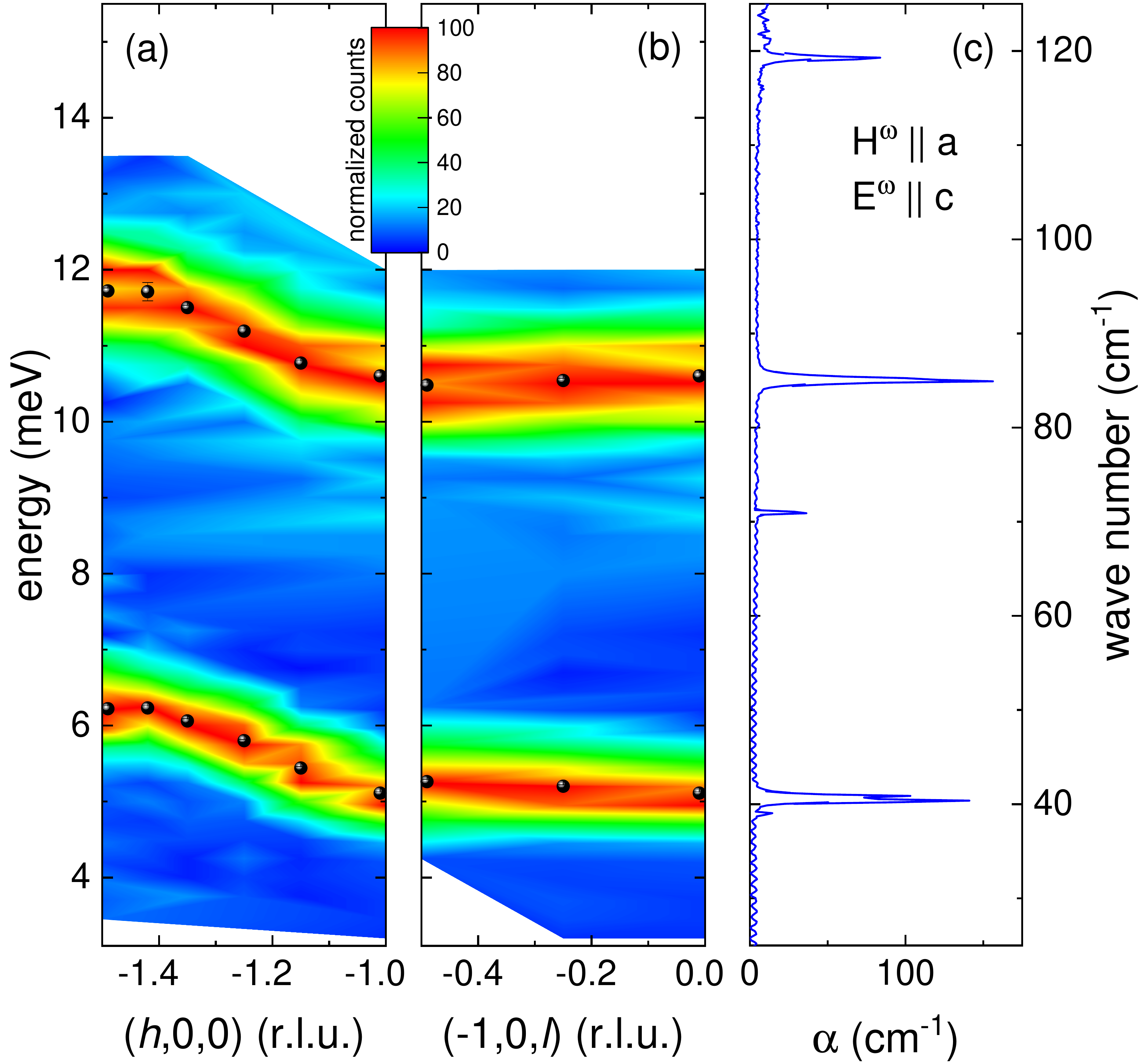}
\caption{\label{fig:0T} (a)/(b) Dispersion of magnon modes along the $(h,0,0)$/$(-1,0,l)$ direction, as determined from inelastic neutron scattering. Symbols indicate the mode energies for the particular energy scans. The colour coding represents the normalized neutron count. (c) Absorption spectrum with absorption coefficient $\alpha$ measured at 2\,K in zero magnetic field for light polarization \Epara{} \& \Hparaa. For direct comparison of the zone-center energies, as obtained by the two techniques, all frames share a common vertical scale, with energy/frequency units displayed on the left/right axis.}
\end{figure}

For the first time in the {$M_{2}$Mo$_3$O$_8$} crystal family (where $M$ stands for a metal ion), the magnon dispersion of {\cmo} was measured in zero magnetic field. The results along the $(h,0,0)$ and $(-1,0,l)$ directions, obtained by energy scans in inelastic neutron scattering, are shown in Fig.~\ref{fig:0T}(a) and (b), respectively. As expected for a four-sublattice antiferromagnetic order, below \TN{} in zero field two intense modes are observed for both directions, which are located around 5 and 10.5\,meV in the zone center. 
Along the $(h,0,0)$-line, both branches show a clear dispersion, while no dispersion can be resolved along the $(-1,0,l)$-line. This indicates that the antiferromagnetic exchange between Co sites in the same honeycomb layer is the dominant magnetic interaction and inter-layer exchange coupling is much weaker, as also found for \mmo\cite{Szaller:2020}.

Figure \ref{fig:0T} provides a direct comparison between the energy of the two magnon modes resolved by inelastic neutron scattering and the energy of the modes observed in the THz absorption spectra in the zone center for light polarization \Epara{} \& \Hparaa{}. The two modes found in the inelastic neutron data show up as the strongest features in the absorption spectrum, centered at 41\,cm$^{-1}\approx 5.1$\,meV and 85\,cm$^{-1}\approx 10.6$\,meV.
However, a closer inspection of the THz spectra reveals additional weaker modes at 71\,cm$^{-1}$ and 119\,cm$^{-1}$. Please note that the considerably lower energy resolution of the neutron scattering experiment does not allow making a statement on the sharp and weak mode at 71\,cm$^{-1}$, while the energy range of the other mode is uncovered by the current neutron study. Concerning the origin of the additional modes observed in the THz measurements, these may be identified either with low-lying transitions of primarily orbital character or with spin-stretching modes present in anisotropic spin systems with $S>1/2$~\cite{Penc:2012}.

\begin{figure*}[h!tb]
\includegraphics[width = 0.9\textwidth]{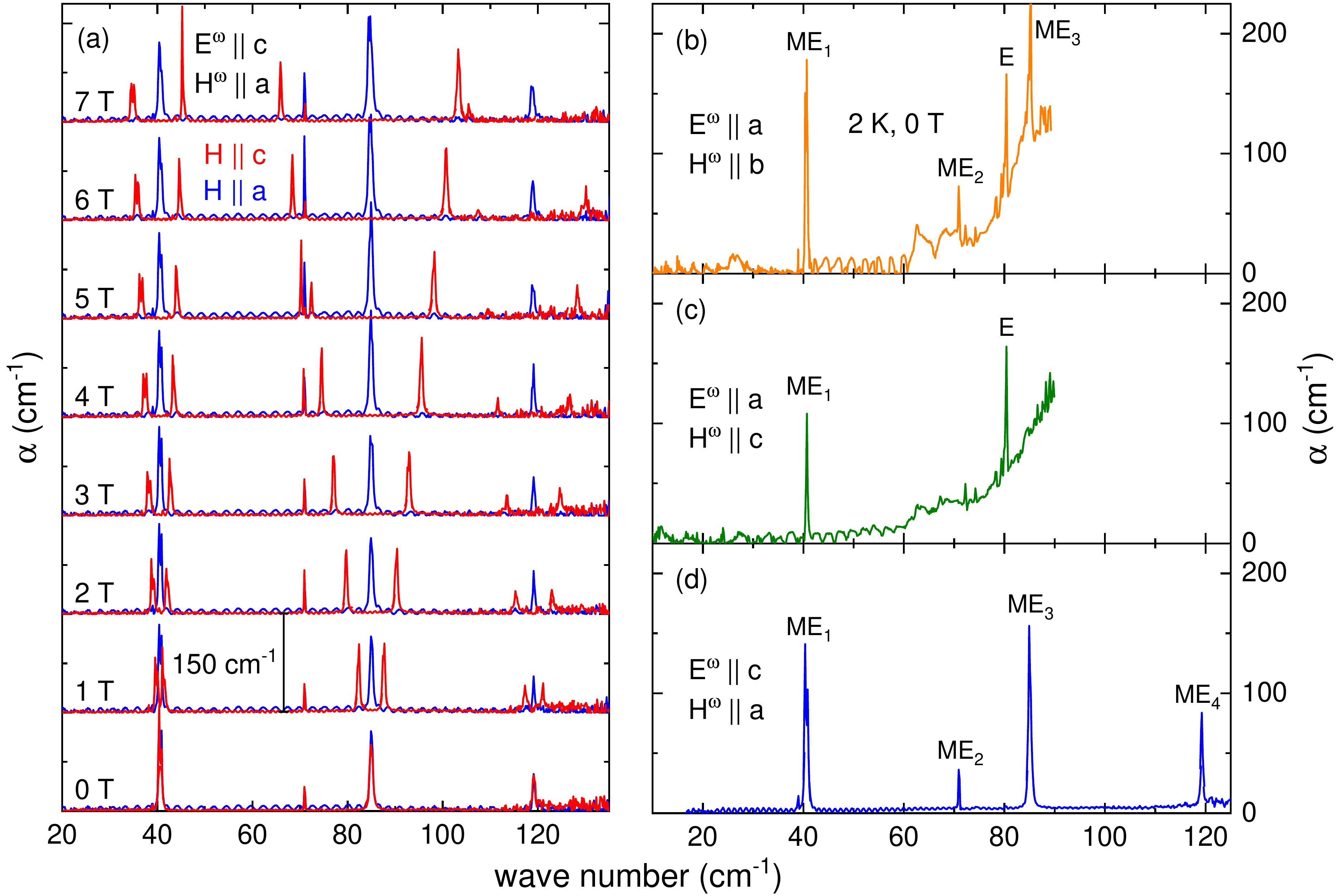}
\caption{\label{fig:waterfall} (a) Absorption coefficient, $\alpha$, spectra measured with light polarization \Epara{} \& \Hparaa{} at 2\,K in magnetic fields $\boldsymbol{H}\parallel c$ (red curves) and $\boldsymbol{H}\parallel a$ (blue curves) up to 7\,T. Spectra are shifted vertically by an offset in proportion to the magnetic field (150\,cm$^{-1}$/T) for clarity. (b)-(d) Zero-field absorption coefficient spectra measured with the three orthogonal light polarizations at 2\,K. Modes are labelled in the same way as in Table 1. }
\end{figure*}

\vspace{10pt}
\noindent
\textbf{Field evolution and selection rules of the modes}\\
The magnetic field dependence of the absorption spectrum is shown in Fig.~\ref{fig:waterfall}(a) for \Epara{} \& \Hparaa{} in static magnetic fields applied parallel (red curves) and perpendicular (blue curves) to the $c$ axis. For $\boldsymbol{H}\parallel c$, all modes except the one at 71\,cm$^{-1}$ show a V-shape splitting with increasing magnetic field.
From the shift of the mode frequencies, which is linearly proportional to the strength of the magnetic field, the effective $g$-factors of the different modes are determined and found to vary over an unusually wide range from 0 to 5.6 (see Table \ref{tab:selectionrules}).
In contrast, for magnetic fields applied perpendicular to the $c$ axis, no splitting of the modes could be observed. This observation implies that the ordered moments have zero projections within the $ab$ plane in the studied magnetic field range, again supporting the easy-axis character of the antiferromagnetic ground state. In addition to measurements performed with polarization \Epara{} \& \Hparaa, the absorption spectra were also studied in the
orthogonal polarization configuration in zero field, as shown in Fig.~\ref{fig:waterfall}(a). The spectra in Fig.~\ref{fig:waterfall}(a) reveal an additional resonance $E$ at 80\,cm$^{-1}$, the strength of which is independent of the orientation of  $\boldsymbol{H}^{\omega}$, but the mode is only active for $\boldsymbol{E}^{\omega}\parallel a$. Such an only electric-dipole active mode is not expected to exhibit directional dichroism~\cite{Kezsmarki:2014}. In contrast, the directional dichroism observed for the other four resonances demonstrates unambiguously that  $ME_1-ME_4$ are magnetoelectric excitations, which are both electric- and magnetic-dipole active. The selection rules as well as the $g$-factors of the modes for $\boldsymbol{H}\parallel c$  are listed in Table \ref{tab:selectionrules}.

\begin{table}[h]
\squeezetable
\begin{ruledtabular}
\centering\footnotesize
   \begin{tabular}{ccccccc}

   \textbf{mode}
	& {\bfseries $\boldsymbol{\omega_0}$} & $g$ &
	
	\multicolumn{3}{c}{\textbf{polarization}} & \textbf{ME activity}
	 \\
    & $[\mathrm{cm}^{-1}]$ &  & $H^{\omega}_b$--$E^{\omega}_a$ & $H^{\omega}_c$--$E^{\omega}_a$
    & \multicolumn{1}{c}
     {$H^{\omega}_a$--$E^{\omega}_c$}  &  \\
    \midrule
	 $ME_1$ & 41 & 1.6& $\checkmark$ & $\checkmark$  &  $\checkmark$
	 &  $H^{\omega}_c$--$E^{\omega}_a$     \\
	 $ME_2$ & 71 & 0.0&  $\checkmark$  & $\times$    &  $\checkmark$
	 &  $H^{\omega}_a$--$E^{\omega}_c$      \\
	 	 $E$ & 80 & 3.3& $\checkmark$ & $\checkmark$    &  $\times$
	 &  --- \\
	 $ME_3$ & 85 &  5.6& $\checkmark$ & $\times$    &  $\checkmark$
	 &  $H^{\omega}_a$--$E^{\omega}_c$   \\
	 $ME_4$ & 119 & 3.7& $\checkmark$ & n.r.    &  $\checkmark$
	 &  $H^{\omega}_a$--$E^{\omega}_c$ \\

    \end{tabular}
    \caption{\label{tab:selectionrules} Excitations observed in \cmo{} below $T_N$. For each mode, we list the resonance frequency $\omega_0$ and the effective $g$-factor for $\boldsymbol{H}\parallel c$, and specify in which polarization configurations the mode is observed/silent ($\checkmark$/$\times$) in zero field. Mode $ME_4$ at 119\,cm$^{-1}$ was not resolved (n.r.) in one polarization configuration because of a low-lying phonon mode with high absorption in the corresponding spectral range. The last column refers to the polarization configuration, where the directional dichroism is observed.}
\end{ruledtabular}
\end{table}

\begin{figure}[tb]
\includegraphics[width = 1\columnwidth]{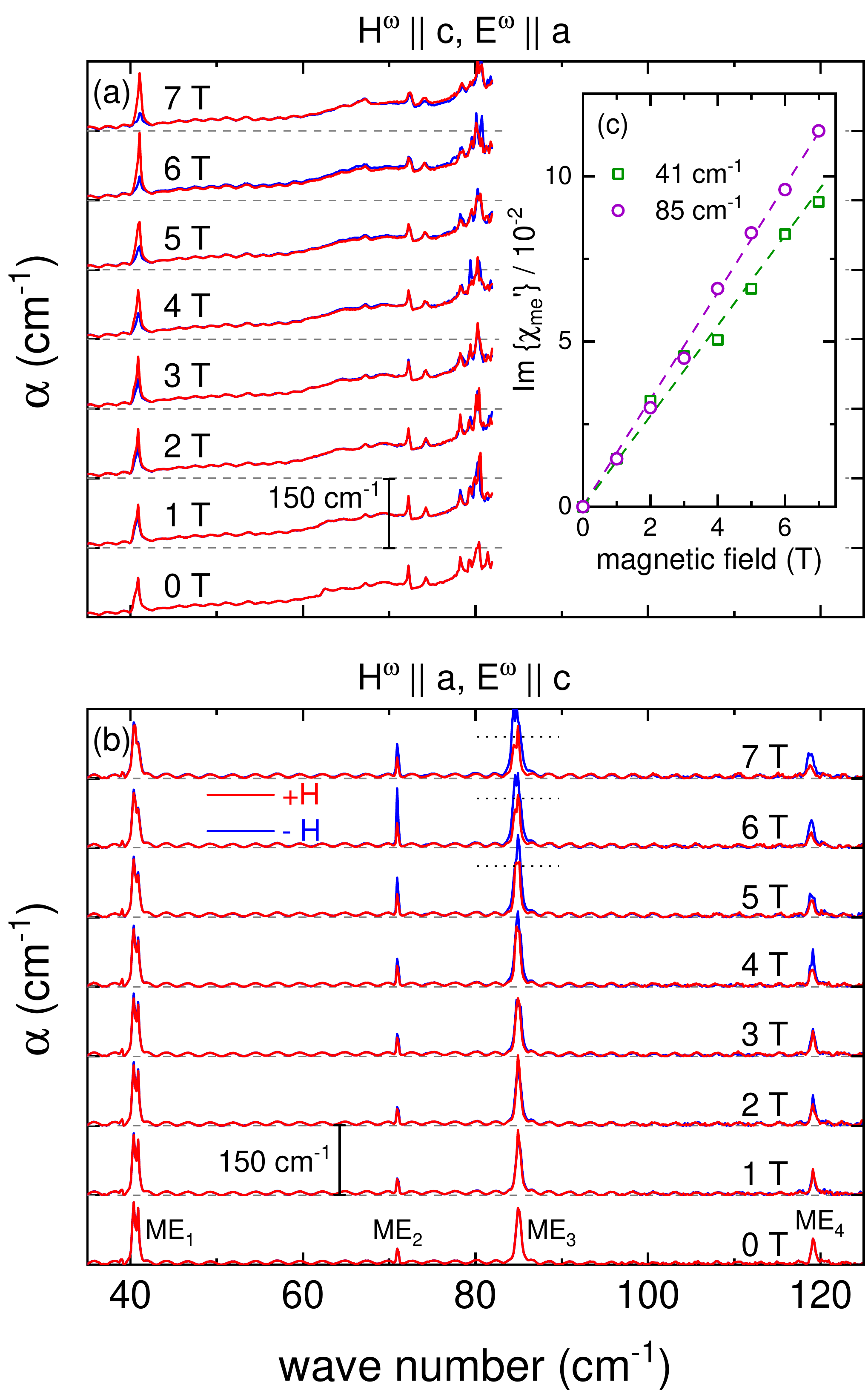}
\caption{\label{fig:DD_fielddep} Magnetic field dependence ($\boldsymbol{H}\parallel a$) of the absorption spectra at 2\,K for polarization (a) \Hpara{} \& \Eparaa{} and (b) \Hparaa{} \& \Epara{}. Red and blue curves correspond to positive and negative magnetic fields, respectively. (c) Magnetic field dependence of the peak value of the imaginary part of $\chi'_\mathrm{me}$ for the modes $ME_1$ and $ME_3$. Note that the maximum of the absorption peak for the latter is not fully resolved due to  its highly increased absorption strength of this mode, as indicated by dashed lines. }
\end{figure}

\begin{figure*}[tb]
\includegraphics[width = 1\textwidth]{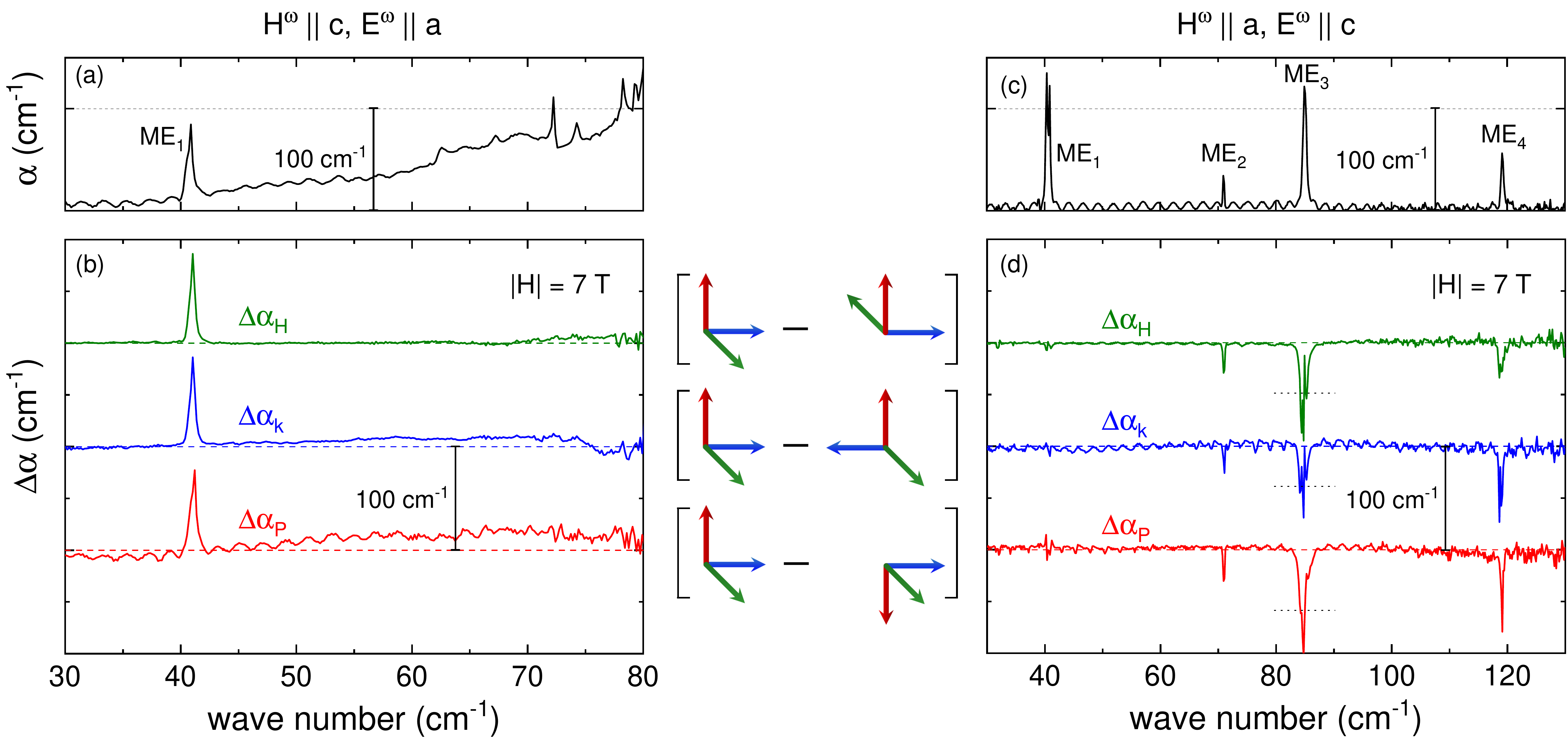}
\caption{\label{fig:DD} (a) Absorption spectra measured at 2\,K and zero magnetic field for polarization \Hpara{} \& \Eparaa{}. (b) Absorption difference spectra $\Delta\alpha_H$, $\Delta\alpha_k$ and $\Delta\alpha_P$ for the same polarization in $|H|=7$\,T$\parallel a$. For clarity, curves are shifted vertically by constant offsets of 100\,cm$^{-1}$. (c),(d) Same set of data for the perpendicular polarization direction \Hparaa{} \& \Epara{} . Dashed lines across the mode at 85\,cm$^{-1}$ indicate that due to the high absorption strength the peak maximum could not be fully resolved. The arrows represent the orientations of $\boldsymbol{k}$, $\boldsymbol{P}$ and $\boldsymbol{M}$ , following the color scheme of Fig. \ref{fig:Crystal_structure}(c).}
\end{figure*}

\vspace{10pt}
\noindent
\textbf{Directional dichroism}\\
According to linear-response theory, the \dd{}
at long wavelengths, i.e. in the THz or far-infrared range covered here, originates from the dynamic magnetoelectric effect~\cite{Szaller:2014}. In the present case, \cmo{} has a spontaneous polarization $\boldsymbol{P}\parallel c$ and by applying the external magnetic field $\boldsymbol{H}\perp c$, we magnetize the material perpendicular to the $c$ axis, $\boldsymbol{M}\parallel a$. This reduces the $6'm'm$ hexagonal magnetic point symmetry of the antiferromagnetic state \cite{Kurumaji:2015} to $2'm'm$ orthorhombic symmetry and activates the time-reversal odd $\chi'_{ac}$ and $\chi'_{ca}$ components in the magnetoelectric tensor \cite{Kezsmarki:2011}.
In this case, the relation
$\delta n\propto\boldsymbol{k}\cdot(\boldsymbol{P}\times\boldsymbol{M})$ predicts directional dichroism for light beams propagating along and opposite to the $b$ axis, being perpendicular to both the $a$ and $c$ axes.
Correspondingly, the solution of the Maxwell equations yields four different values for the refractive index $n$, two for each of the two orthogonal linear polarization configurations \cite{Kezsmarki:2011}:
\begin{eqnarray}
n^{\pm k_b}_{ac}&\approx& \sqrt{\epsilon'_{aa}\mu'_{cc}}\mp\chi'_{ca},\label{ref_index12} \\
n^{\pm k_b}_{ca}&\approx& \sqrt{\epsilon'_{cc}\mu'_{aa}}\pm\chi'_{ac}.
\label{ref_index34}
\end{eqnarray}
Here $n^{\pm k_b}_{ac}$ and $n^{\pm k_b}_{ca}$ denote the refractive indices for light polarizations \Eparaa{} \& \Hpara{} and \Epara{} \& \Hparaa{}, respectively. States propagating in opposite directions are indicated by $\pm k_b$. $\epsilon'_{ij}$ and $\mu'_{ij}$ are time-reversal invariant components of the electric permittivity and magnetic permeability tensors, respectively. In the following, we discuss the directional dichroism in these two configurations and sequentially reverse $\boldsymbol{k}$, $\boldsymbol{P}$ and $\boldsymbol{M}$, in order to prove the trilinear form of the optical magnetoelectric effect.

Figures~\ref{fig:DD_fielddep}(a) and (b) show the absorption spectra at 2\,K for the two polarization configurations measured in positive (red) and negative (blue) magnetic fields $\boldsymbol{H}\parallel a$. 
For the polarization \Eparaa{} \& \Hpara{}, \dd{} shows up for the $ME_1$ mode at 41\,cm$^{-1}$ at finite magnetic fields. The absorption strength varies linearly with magnetic field (or the induced transverse magnetization), i.e. it is enhanced and reduced for positive and negative magnetic fields, respectively. Notably, in -7\,T the absorption of this resonance is almost fully suppressed and one-way transparency is nearly achieved, as will be directly evidenced in the following by the reversal of the propagation direction.

For the orthogonal light polarization \Epara{} \& \Hparaa{}, the absorption strength of the $ME_1$ mode at 41\,cm$^{-1}$ is independent of the magnetic field, while for the modes $ME_2$, $ME_3$, $ME_4$ at 71, 85 and 119\,cm$^{-1}$, respectively, the absorption coefficient shows a field dependence. However, in this case, it is reduced for positive and enhanced for negative fields. According to Eqs.~\ref{ref_index12} and \ref{ref_index34}, this leads to the same sign of $\chi'_{ca}$ and $\chi'_{ac}$, generating the \dd{} in the first and the second polarization configuration, respectively.

The spectra of the magnetoelectric coefficients can be calculated from the absorption difference between oppositely magnetized states of the material according to
\begin{eqnarray}
    &&\Im\{\chi\prime(\omega)\} = \frac{c}{4\omega}\Delta\alpha_M(\omega)\nonumber\\&&=\frac{c}{4\omega}\{\alpha(\omega)[+\boldsymbol{k},+\boldsymbol{P},+\boldsymbol{M}]-\alpha(\omega)[+\boldsymbol{k},+\boldsymbol{P},-\boldsymbol{M}]\}. \nonumber
\end{eqnarray}

The imaginary part of the dimensionless magnetoelectric coefficient spectrum $\chi'_{ca}$ has a well-resolved maximum at 41\,cm$^{-1}$, which follows a linear field dependence, as in the inset of Fig.~\ref{fig:DD_fielddep}(a). For the mode at 85\,cm$^{-1}$, the absorption peak is not properly resolved due to the high absorption, thus the foot of this absorption peak was fitted by a Lorentzian for both field directions to estimate the value of $\chi'_{ac}$ at the resonance. The corresponding values of $\Im\{\chi'_{ac}$\}, also plotted in the inset of Fig.~\ref{fig:DD_fielddep}(a), follow a linear field dependence similar to $\Im\{\chi'_{ca}$\}.

In order to prove that the term in the refractive index is also an odd function of $\boldsymbol{P}$ as well as $\boldsymbol{k}$, we study the absorption coefficient upon the reversal of either the polarization of the material or the light propagation direction. Since the material is pyroelectric, the polarization $\boldsymbol{P}$ cannot be reversed by applying an electric field, instead by rotating the sample  by 180$^\circ$ about the $a$ axis. The reversal of the propagation direction $\boldsymbol{k}$ is conveniently done, without any rearrangement of the rest of the optical path, by exchanging THz emitter and receiver in our fiber-coupled Teraflash device. Both of these operations can be performed independently of each other and of the sign change of the magnetic field.

The three absorption difference spectra $\Delta\alpha_M(\omega)$, $\Delta\alpha_P(\omega)$ and $\Delta\alpha_k(\omega)$ are calculated based on experiments carried out at 2\,K with  $\boldsymbol{H}\parallel a$ and $\mu_0|H|=7$\,T  according to
\begin{eqnarray}
\Delta\alpha_M(\omega) &=& \alpha(\omega)[+\boldsymbol{k},+\boldsymbol{P},+\boldsymbol{M}]- \alpha(\omega)[+\boldsymbol{k},+\boldsymbol{P},-\boldsymbol{M}], \nonumber \\                    \Delta\alpha_P(\omega) &=& \alpha(\omega)[+\boldsymbol{k},+\boldsymbol{P},+\boldsymbol{M}]- \alpha(\omega)[+\boldsymbol{k},-\boldsymbol{P},+\boldsymbol{M}], \nonumber \\
\Delta\alpha_k(\omega) &=& \alpha(\omega)[+\boldsymbol{k},+\boldsymbol{P},+\boldsymbol{M}]- \alpha(\omega)[-\boldsymbol{k},+\boldsymbol{P},+\boldsymbol{M}]. \nonumber
\end{eqnarray}
The corresponding spectra are shown in Fig.~\ref{fig:DD} for both polarization directions. The inset in the middle of the figure is a graphical illustration of how the three absorption differences are calculated according to the three equations above.

The three absorption difference spectra $\Delta\alpha_M$, $\Delta\alpha_P$ and $\Delta\alpha_k$ nicely coincide, as expected if the refractive index contains a term $\delta n\propto\boldsymbol{k}\cdot(\boldsymbol{P}\times\boldsymbol{M})$. Remarkably, for the polarization \Eparaa{} \& \Hpara{} the absorption difference in 7\,T, displayed in Fig.~\ref{fig:DD}(b) , is larger than the total absorption in zero field [Fig.~\ref{fig:DD}(a)] for the $ME_1$ mode. In the \Epara{} \& \Hparaa{} polarization configuration, the absorption difference in 7\,T, plotted in Fig.~\ref{fig:DD}(d), for $ME_2$, $ME_3$ and $ME_4$ is close to or larger than the absorption in zero field [Fig.~\ref{fig:DD}(c)]. Specific to $ME_3$, the absorption peak is too strong to be fully resolved in 7\,T. By fitting the flanks of $\Delta\alpha$ spectra with a Lorentzian line (not shown here) yields the maximum absorption difference $\Delta\alpha\approx 200$\,cm$^{-1}$. This value is nearly twice as large as the absorption in zero field, indicating that one-way transparency \cite{Kezsmarki:2014} is realized for this mode.

\section*{Conclusions}
By independent reversal of magnetization, polarization and light propagation, all three possibilities of realizing \dd{} have been demonstrated for the first time. Our THz spectroscopy results on \cmo{} with orthogonal magnetization and polarization directly confirm the trilinear-product form of the optical magnetoelectric effect $\boldsymbol{k}\cdot(\boldsymbol{P}\times\boldsymbol{M})$, which was expected on theoretical grounds, but not confirmed yet in magnetoelectric crystals on a purely experimental basis. Some of the low-energy excitations of this compound, presumably spin-waves, exhibit strong directional dichroism close to the one-way-transparency limit.

\section*{Methods}
\noindent
\textbf{Synthesis}\\
Single crystals were grown by the chemical transport reaction method at temperatures between 900--950\,$^\circ$C using anhydrous TeCl$_4$ as the source of the transport agent. X-ray analysis of the crushed single crystals revealed a single phase composition with a hexagonal symmetry and space group $P6_3mc$. The room-temperature lattice constants $a = b = 5.7677(1)$\,\AA{}  and $c = 9.9097(2)$\,\AA{}  are close to the data reported in literature \cite{Bertrand:1975,Abe:2010}. Single crystals were characterized by magnetometry and specific heat measurements, which confirmed the onset of long-range antiferromagnetic order at $\TN{}=40$\,K.

\noindent
\textbf{THz spectroscopy}\\
Temperature and magnetic field dependent time-domain THz spectroscopy measurements were performed on plane parallel $ab$- and $ac$-cut single crystals of \cmo{} in transmission configuration. For the optical measurements a Toptica TeraFlash time-domain THz spectrometer was used in combination with a superconducting magnet, which allows for measurements at temperatures down to 2\,K and in magnetic fields up to $\pm 7$\,T.

\noindent
\textbf{Inelastic neutron scattering}\\
Inelastic neutron scattering (INS) was measured on two co-aligned crystals with a total weight of $800$\,mg using the thermal neutron triple-axis spectrometer EIGER at the SINQ, Paul Scherrer Institut, Switzerland.
The use of a double focusing PG(002) monochromator and analyzer gave the energy resolution $0.7$\,meV at the elastic line. The final wave vector $k_\mathrm{f} =2.66$\,\AA$^{-1}$ was filtered by a PG filter. The sample was mounted in a ILL cryostat with the $ac$-plane horizontal, which gave access to excitations along the principal directions $(h,0,0)$ and $(-1,0,l)$.

\section*{Data Availability}
The data that support the findings of this study are available from the
corresponding author upon reasonable request.

\begin{acknowledgements}
J.D. acknowledges stimulating discussions with Prof. Jorge Stephany. This research was partly funded by Deutsche Forschungsgemeinschaft DFG via the Transregional Collaborative Research Center TRR 80 “From Electronic correlations to functionality” (Augsburg, Munich, Stuttgart).  The support via the project ANCD 20.80009.5007.19 (Moldova) is also acknowledged. This research was supported by the National Research, Development and Innovation Office – NKFIH, FK 153003 and Bolyai 00318/20/11. D.S. acknowledges the support of the Austrian Science Fund (FWF) [I 2816-N27, TAI 334-N] and that of the Austrian Agency for International Cooperation in Education and Research [WTZ HU 08/2020].
\end{acknowledgements}

\section*{Author Contributions}
L.P. and V.T. synthetized and characterized the crystals; S.R., D.G.F., and A.S. performed the THz measurements; S.R., J.D., and I.K. analyzed the THz measurements; S.G., K.G., O.Z. performed and analyzed the neutron scattering experiments. S.K, J.D, and I.K. wrote the manuscript with contributions from D.S. and S.B.; I.K. planned the project.

\section*{Competing Interests}
The authors declare that there are no competing interests.

\end{document}